# Understanding Energy-Level Structure Using a Quantum Rubik's Cube

Yu Wang, Maolin Bo*

Key Laboratory of Extraordinary Bond Engineering and Advanced Materials Technology (EBEAM) of Chongqing, Yangtze Normal University, Chongqing 408100, China

*Corresponding author: Maolin Bo (E-mail addresses: bmlwd@yznu.edu.cn)

Abstract

This study combines the quantum Rubik's Cube matrix with the Benalcazar–Bernevig–Hughes model, defines a matrix algorithm based on the reverse process of convolution, and constructs an expression for the quantum Rubik's Cube matrix and Hamiltonian. Furthermore, in order to make the operation of the quantum Rubik's Cube matrix clearer, we use a Josephus ring to draw a topological graph of the Rubik's Cube expansion. This article uses a quantum Rubik's Cube to calculate energy-level transitions of electrons, and shows that its operation corresponds to path integration. The band dispersion is obtained. This work provides new ideas and methods for calculating Hamiltonians and studying energy-level structure.

## 1. Introduction

Heisenberg established the matrix form of quantum mechanics and linked the Hamiltonian of quantum mechanics with matrices[1-5]. The Hamiltonians of matrices with different mathematical structures handle different scientific problems. Therefore, the various mathematical expressions of Hamiltonians are crucial for solving physical problems of different origins. The most common example of gauge field theory for non-abelian symmetric groups (also known as non-abelian groups) [6-12] is the Yang–Mills theory. Physical systems are often expressed as Lagrangians invariant under certain transformations, and when transformations are performed simultaneously at each spatiotemporal point, they have global symmetry. Yang and Mills described the physics of elementary-particle interactions, describing elementary particles as fields and including the strong and electromagnetic interactions[13-18]. In strong interactions, the Yang–Mills field has non-abelian group symmetry, so its gauge transformation cannot be expressed simply as a complex number multiplied by a phase factor. We have used the rotation of a quantum Rubik's Cube[19] to represent Pauli matrix multiplication in the Ising model[20, 21]. The quantum Rubik's Cube generates a path, which is the propagator in path integration. Path changes can drive wave-function changes; this is equivalent to manipulating the wave function.

Group theory is an important tool and theoretical framework used in quantum mechanics to describe symmetry and dynamic behavior in physical systems. Non-abelian symmetry also holds a significant position in quantum physics. The term "non-abelian" means that the symmetry does not satisfy the properties of abelian groups: the order of symmetry operations is important, and different commutation orders may lead to different results. This study is based on the non-abelian group governing the Rubik's Cube matrix. We utilized the rotation operation of the quantum Rubik's Cube matrix with non-abelian group characteristics and modeled its structure using the Benalcazar–Bernevig–Hughes (BBH) model[22]. We constructed the Hamiltonian of the fourth-order quantum Rubik's Cube matrix, solved the eigenvalues of the Hamiltonian, and plotted phase images obtained from non-abelian group operations in complex *k*-space.

This study uses the Josephus ring[23] to draw a topological diagram that reduces the quantum Rubik's Cube from three dimensions to two for restoration processing, and obtains its topological diagram in the form of energy-level transitions. Finally, based on perturbation theory, a fifth-order quantum Rubik's Cube matrix is used to simulate the transition of electrons between the *s* and *p* orbitals. The mathematical structure is equivalent to the principle of the path-integral propagator.

## 2. Results and discussion

### 2.1 Hamiltonian Analysis of Fourth-Order Quantum Rubik's Cube

The Hamiltonian of a single quantum Rubik's Cube model can be expressed as[24]:

$$H = \pm \sum_j J_j [\![ M ]\!]_{i\times i} , \tag{1}$$

where $J_i$ is the coupling function and $[\![ M ]\!]_{i\times i}$ is a Rubik's (magic) cube matrix. The index $i$ ($>1$) represents the order (size) of the Rubik's Cube matrix; $j(\leq 6)$ represents the number of Rubik's Cube matrices. The $j$ of a strange Rubik's Cube matrix can be > 6. The sign of $J$ is the sign of the coupling. Furthermore, the expression for the complex multi-body quantum Rubik's Cube Hamiltonian can be written as

$$H = \pm \sum_{j,k} J_{jk} [\![ M_j ]\!]_{i\times i} , \tag{2}$$

where $k(>1)$ indicates the number of quantum Rubik's Cubes. In this study, we connect the quantum Rubik's Cube with the Ising model to construct a new Hamiltonian mathematical form. The operational rules for a quantum Rubik's Cube can be found in **Support Information**.[25, 26] They satisfy the allocation rate, exchange rate, and associative law of addition. Furthermore, the matrix of the quantum Rubik's Cube can be convolved, providing the possibility for quantum computing in neural networks.

Using the quantum Rubik's cube matrix $[\![ M ]\!]_{i\times i}$ as the input $X^i$ of the convolutional neuron model, in order to calculate the output feature map $Y^p$, convolution kernels $W^{p,1}$, $W^{p,2}$,…, $W^{p,D}$ are used to convolution the input feature maps $X^1$, $X^2$,…, $X^D$, respectively. Then, the convolution results are added and a scalar bias $b^p$ is added to obtain the net input $Z^p$ of the convolutional layer. After passing through a nonlinear activation function, the output feature map $Y^p$ is obtained.

$$\begin{cases} Z^p = W^p \otimes [\![ M ]\!] + b^p = \sum_{d=1}^{D} W^{p\,d} \otimes [\![ M ]\!]^d + b^p \\ Y^p = f\left(Z^p\right) \end{cases} \tag{3}$$

Among them, $W^p \in \mathbb{R}^{U\times V\times D}$ is the three-dimensional convolution kernel, $f(\cdot)$ f is the nonlinear activation function, and the ReLU function is generally used.[27]

The entire calculation process is shown in **Fig.1**. To enable the convolutional layer to output $P$ feature maps, the above calculation process can be repeated $P$ times to obtain $P$ output feature maps $Y^1$, $Y^2$,…, $Y^p$. In the convolutional layers with input $\chi \in \mathbb{R}^{M\times N\times D}$ and output $\gamma \in \mathbb{R}^{M'\times N'\times P}$, each output feature map requires $D$ convolution kernels and a bias.

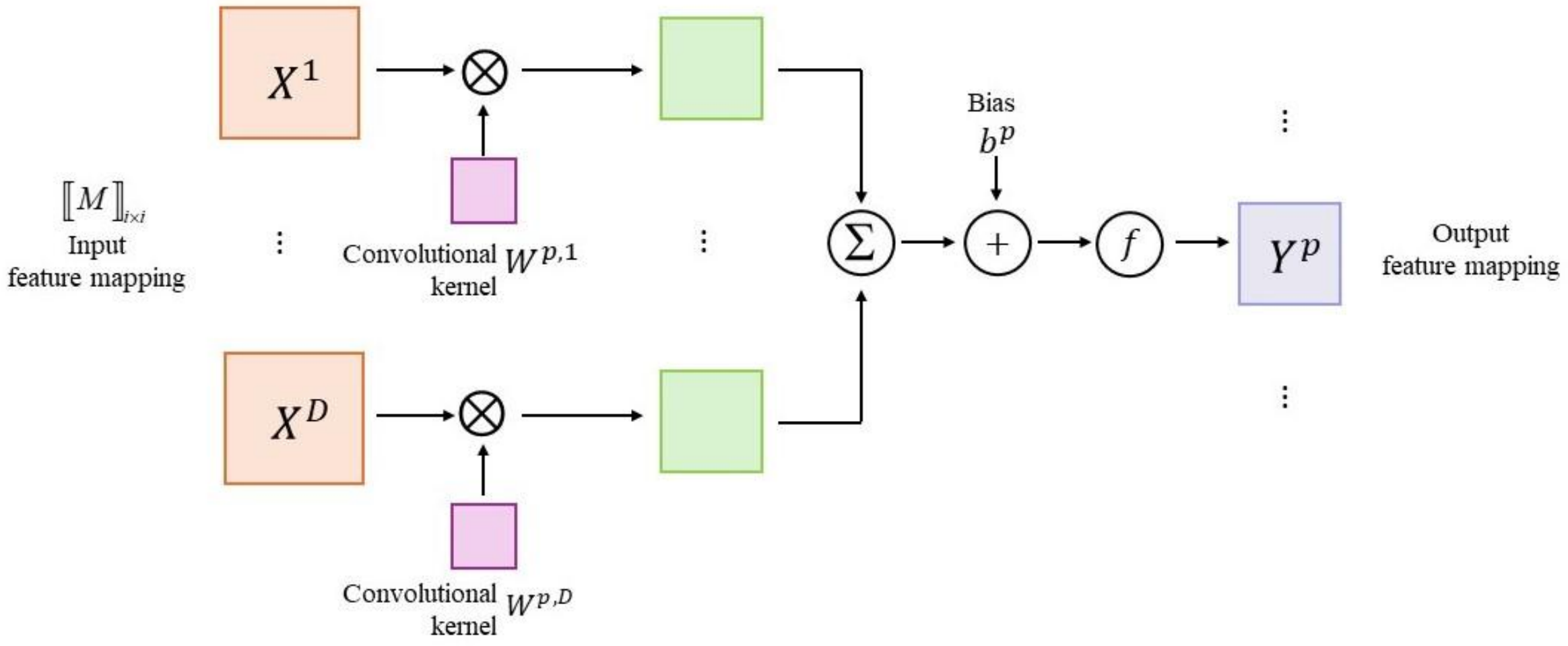

**FIG. 1** Input Feature Mapping Group $X$ to Output Feature Mapping $Y^p$ in Convolutional Layer

For the BBH model [22], we use the matrix expansion of the fourth-order quantum Rubik's Cube to calculate the Hamiltonian:

$$\begin{cases} H(k,\delta)=\sum_j J_j [\![M]\!]_{i\times i} = (\gamma+\lambda\cos(k_x))\Gamma_4+\lambda\sin(k_x)\Gamma_3+(\gamma+\lambda\cos(k_y))\Gamma_2+\lambda\sin(k_y)\Gamma_1+\delta\Gamma_0 \\ \qquad\qquad (\Gamma_4) \\ [\![M]\!]_{4\times 4}=(\Gamma_3)(\Gamma_0)(\Gamma_1)(\Gamma_5) \\ \qquad\qquad (\Gamma_2) \end{cases} \tag{3}$$

Here, $J_0=\delta$, $J_1=\lambda\sin(k_y)$, $J_2=\gamma+\lambda\cos(k_y)$, $J_3=\lambda\sin(k_x)$, $J_4=\gamma+\lambda\cos(k_x)$, $J_5=0$. The Rubik's Cube matrices are $\Gamma_0=\tau_3\sigma_0$, $\Gamma_k=\tau_2\sigma_k$ and $\Gamma_4=\tau_1\sigma_0$, for $k$ = 1, 2, and 3; $\tau$, $\sigma$ are Pauli matrices for the degrees of freedom within a unit cell. Therefore,

$$\Gamma_0=\tau_3\sigma_0=\begin{vmatrix} 1 & 0 & & \\ 0 & 1 & & \\ & & -1 & 0 \\ & & 0 & -1 \end{vmatrix},\quad \Gamma_1=-\tau_2\sigma_1=-\begin{vmatrix} & & 0 & -i \\ & & -i & 0 \\ 0 & i & & \\ i & 0 & & \end{vmatrix}$$

$$\Gamma_2=-\tau_2\sigma_2=-\begin{vmatrix} & & 0 & -1 \\ & & 1 & 0 \\ 0 & 1 & & \\ -1 & 0 & & \end{vmatrix},\quad \Gamma_3=-\tau_2\sigma_3=-\begin{vmatrix} & & -i & 0 \\ & & 0 & i \\ i & 0 & & \\ 0 & -i & & \end{vmatrix},\quad \Gamma_4=\tau_1\sigma_0=\begin{vmatrix} & & 1 & 0 \\ & & 0 & 1 \\ 1 & 0 & & \\ 0 & 1 & & \end{vmatrix}$$

$$\text{and}\quad \Gamma_5=\begin{vmatrix} & & 0 & 0 \\ & & 0 & 0 \\ 0 & 0 & & \\ 0 & 0 & & \end{vmatrix}. \tag{4}$$

The absolute value of the transition term inside the cell is $\gamma$, and the absolute value of the transition term between cells is $\lambda$.

**Fig. 2** shows the operation and topology of the quantum Rubik's Cube matrix. **Fig. 2 (a)** shows the operation of a fourth-order quantum Rubik's Cube. $U_{m\,(n)}$, $R_{m\,(n)}$, $F_{m(n)}$, $B_{m\,(n)}$, $TR_{m\,(n)}$, $TU_{m\,(n)}$, $TL_{m\,(n)}$, $MU_{m\,(n)}$ and $MR_{m\,(n)}$ are rotation operators of the quantum Rubik's Cube; $F'_{m(n)}$, $U'_{m\,(n)}$, $R'_{m\,(n)}$, $F'_{m(n)}$, $B'_{m\,(n)}$, $TR'_{m\,(n)}$, $TU'_{m\,(n)}$ and $TL'_{m\,(n)}$ are the rotation operators of its inverse. The group of the quantum Rubik's Cube is defined as $\Re$ = { $U_{m\,(n)}$, $R_{m\,(n)}$, $F_{m(n)}$, $B_{m\,(n)}$, $TR_{m\,(n)}$, $TU_{m\,(n)}$, $TL_{m\,(n)}$, $MU_{m\,(n)}$, $MR_{m\,(n)}$, $F'_{m(n)}$, $U'_{m\,(n)}$, $R'_{m\,(n)}$, $F'_{m(n)}$, $B'_{m\,(n)}$, $TR'_{m\,(n)}$, $TU'_{m\,(n)}$, $TL'_{m\,(n)}$ }. A letter with no subscripts (e.g., F) is an instruction to turn that face 90° clockwise with respect to the center of the cube. A letter with an apostrophe (F') denotes a 90° counter-clockwise turn. A letter followed by the subscript 2 ($F_2$) denotes two turns, i.e., a 180° turn. The subscript *n* indicates the number of operations, and *m* represents the steps of each operation. The fifth-order quantum Rubik's Cube is described similarly.

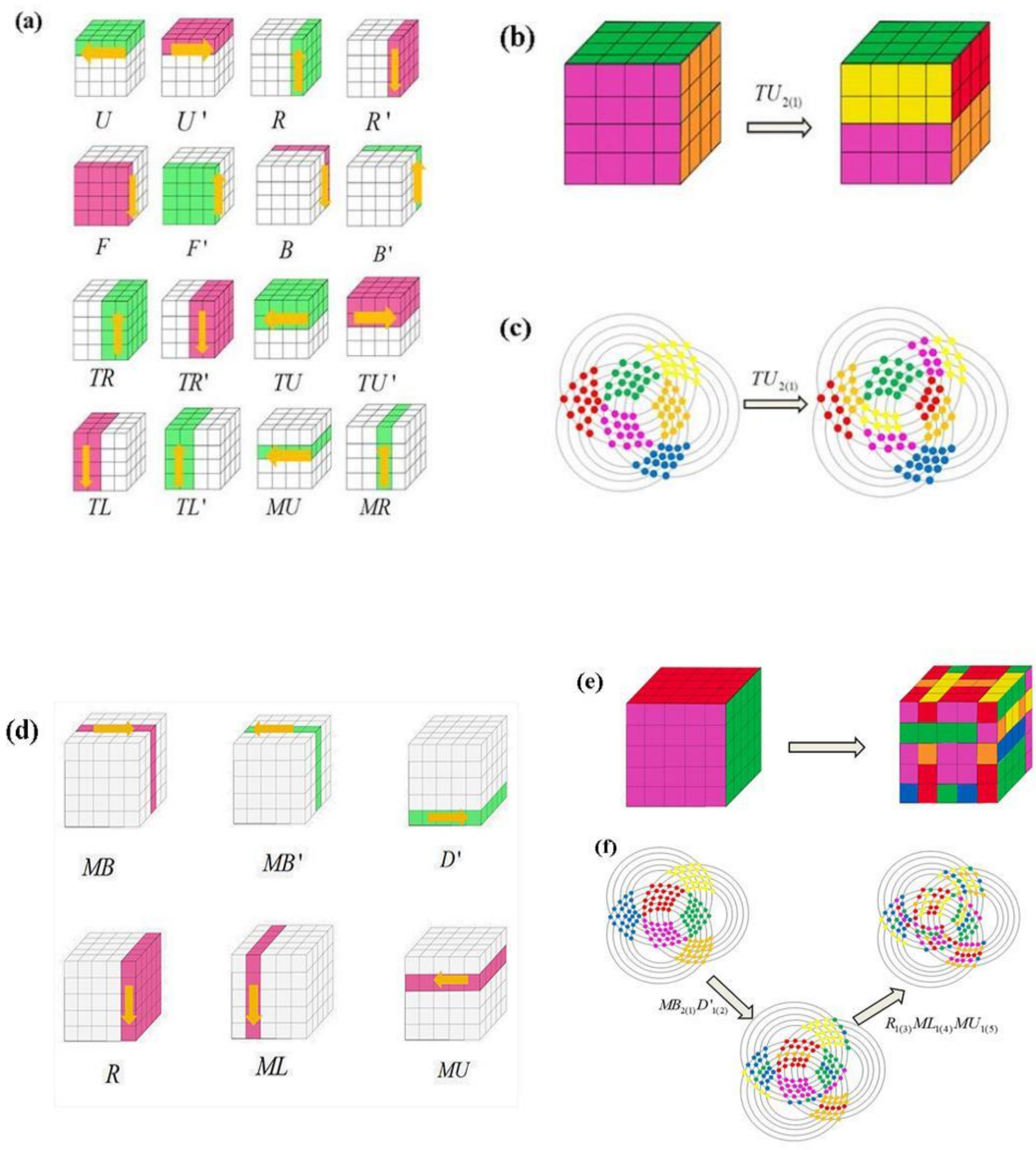

**FIG. 2 (a) (d)** Rules for operating the fourth- and fifth-order quantum Rubik's Cube. **(b) (e)** The initial and final states of the fourth- and fifth-order Rubik's Cubes. **(c) (f)** Topological expansion diagram using Josephus rings to represent fourth- and fifth-order Rubik's Cubes.

A Josephus ring is usually solved by a looped linked list approach, which uses a structured array to form a looped chain. There are two points in the structure, one of which is represented as a pointer to the next point, and the other is a marker indicating whether this point should be eliminated. Applying a Josephus ring can effectively transform the corresponding positions of Rubik's Cube.

The topological structure of Rubik's Cube is formed by the intersection of three sets of Josephus rings, with a total of six sets of intersections. The area where the pink dots are located in the Josephus ring on the left side of **Fig. 2 (c)(f)** is always $\Gamma_0$ of the Rubik's Cube. Similarly, the orange area is $\Gamma_1$, the blue area is $\Gamma_2$, the red area is $\Gamma_3$, the green area is $\Gamma_4$, and the yellow area is $\Gamma_5$. In **Fig. 2 (c)**, four rings with the same center but different radii are shown as a set of concentric circles. Josephus rings are used to represent each layer of the Rubik's Cube, showing the two-dimensional expansion of the Rubik's Cube after rotating according to the rotation rules in its initial state. Each intersection represents a square on the Rubik's Cube, equivalent to one of the six faces of the cube, with 4×4 points on each face. Rotating a face around a certain axis is equivalent to exchanging two sets of Josephus rings.

The BBH model combines Ising model with the matrix operation Hamiltonian of quantum Rubik's Cube. We use the matrix of a fourth-order quantum Rubik's Cube to perform dynamic simulations on the constructed Hamiltonian:

$$\begin{cases} H(k,\delta)=(\gamma+\lambda\cos(k_x))\Gamma'_4+\lambda\sin(k_x)\Gamma'_3+(\gamma+\lambda\cos(k_y))\Gamma'_2+\lambda\sin(k_y)\Gamma'_1+\delta\Gamma'_0 \\ [\![M]\!]_{4\times4}\left|TU_{2(1)}\right\rangle=[\![M']\!]_{4\times4} \\ \begin{array}{ll} & \qquad\quad (\Gamma'_4) \\ [\![M']\!]_{4\times4}= & (\Gamma'_3)(\Gamma'_0)(\Gamma'_1)(\Gamma'_5) \\ & \qquad\quad (\Gamma'_2) \end{array} \end{cases} \tag{5}$$

Here,

$$\Gamma'_0=\tau_3\sigma_0=\begin{vmatrix} 0 & 0 & & \\ 0 & 0 & & \\ & & -1 & 0 \\ & & 0 & -1 \end{vmatrix},\ \Gamma'_1=\begin{vmatrix} & & i & 0 \\ & & 0 & -i \\ 0 & -i & & \\ -i & 0 & & \end{vmatrix},\ \Gamma'_2=\begin{vmatrix} & & 0 & 1 \\ & & -1 & 0 \\ 0 & -1 & & \\ 1 & 0 & & \end{vmatrix},$$

$$\Gamma'_3=\begin{vmatrix} & & 0 & i \\ & & i & 0 \\ -i & 0 & & \\ 0 & i & & \end{vmatrix},\ \Gamma'_4=\begin{vmatrix} & & 1 & 0 \\ & & 0 & 1 \\ 1 & 0 & & \\ 0 & 1 & & \end{vmatrix},\ \text{and}\ \Gamma'_5=\begin{vmatrix} 1 & 0 & & \\ 0 & 1 & & \\ & & 0 & 0 \\ & & 0 & 0 \end{vmatrix}. \tag{6}$$

In the fourth-order quantum Rubik's Cube, the TU are groups $\Re$ that operate on the fourth-order Rubik's Cube matrix. The $\left|TU\right\rangle$ are group operators that operate on the quantum Rubik's Cube; $\left|TU_{2(1)}\right\rangle$ indicates two revolutions to the left on the top two layers of the Rubik's Cube. Therefore, the operation of a quantum Rubik's Cube can be written in the following form:

$$H|\Re\rangle = \pm\sum_{j,k} J_{jk} [\![ M_j ]\!]_{i\times i} |\Re\rangle = \pm\sum_{j,k} J_{jk} [\![ M'_j ]\!]_{i\times i} .$$

Although we used a concise Dirac symbol to represent this quantum Rubik's Cube matrix operation, the reality is different. The quantum Rubik's Cube matrix operator is a very special type of propagation operator that only represents the operation of matrix rows and columns. It is equivalent to a path-integral propagator (see **Support Information**).

According to **Eq. 6** , the Hamiltonian expansion after operation is written as:

$$H(k,\delta) = (\gamma + \lambda\cos(k_x))\Gamma'_4 + \lambda\sin(k_x)\Gamma'_3 + (\gamma + \lambda\cos(k_y))\Gamma'_2 + \lambda\sin(k_y)\Gamma'_1 + \delta\Gamma'_0 ,$$

$$H(k,\delta) = \begin{vmatrix} 0 & 0 & i\lambda\sin(k_y) & 2\gamma + \lambda\left(\cos(k_x)+\cos(k_y)\right) + i\lambda\sin(k_x) \\ 0 & 0 & -2\gamma - \lambda\left(\cos(k_x)+\cos(k_y)\right) + i\lambda\sin(k_x) & -i\lambda\sin(k_y) \\ \gamma + \lambda\cos(k_x) - i\lambda\sin(k_x) & -\gamma - \lambda\cos(k_y) - i\lambda\sin(k_y) & -\delta & 0 \\ \gamma + \lambda\cos(k_y) - i\lambda\sin(k_y) & \gamma + \lambda\cos(k_x) + i\lambda\sin(k_x) & 0 & -\delta \end{vmatrix}$$

$$= \begin{vmatrix} 0 & 0 & i\lambda\sin(k_y) & 2\gamma + \lambda\exp(ik_x) + \cos(k_y) \\ 0 & 0 & -2\gamma - \lambda\exp(-ik_x) - \lambda\cos(k_y) & -i\lambda\sin(k_y) \\ \gamma + \lambda\exp(-ik_x) & -\gamma - \lambda\exp(ik_y) & -\delta & 0 \\ \gamma + \lambda\exp(-ik_y) & \gamma + \lambda\exp(ik_x) & 0 & -\delta \end{vmatrix}$$

(7)

We operate on the Hamiltonian matrix with quantum Rubik's Cubes. Numerical calculation with $(\gamma = 1, \lambda = 1, x = y)$ results in the eigenvalues:

$$-\mathrm{e}^{-\mathrm{i}y}\sqrt{2\mathrm{e}^{\mathrm{i}y} + 4\mathrm{e}^{2\mathrm{i}y} + 2\mathrm{e}^{3\mathrm{i}y} - \sqrt{-\,\mathrm{e}^{\mathrm{i}y}(1+\mathrm{e}^{\mathrm{i}y})^6}},$$

$$\mathrm{e}^{-\mathrm{i}y}\sqrt{2\mathrm{e}^{\mathrm{i}y} + 4\mathrm{e}^{2\mathrm{i}y} + 2\mathrm{e}^{3\mathrm{i}y} - \sqrt{-\,\mathrm{e}^{\mathrm{i}y}(1+\mathrm{e}^{\mathrm{i}y})^6}},$$

$$-\mathrm{e}^{-\mathrm{i}y}\sqrt{2\mathrm{e}^{\mathrm{i}y} + 4\mathrm{e}^{2\mathrm{i}y} + 2\mathrm{e}^{3\mathrm{i}y} + \sqrt{-\,\mathrm{e}^{\mathrm{i}y}(1+\mathrm{e}^{\mathrm{i}y})^6}},$$

$$\mathrm{e}^{-\mathrm{i}y}\sqrt{2\mathrm{e}^{\mathrm{i}y} + 4\mathrm{e}^{2\mathrm{i}y} + 2\mathrm{e}^{3\mathrm{i}y} + \sqrt{-\,\mathrm{e}^{\mathrm{i}y}(1+\mathrm{e}^{\mathrm{i}y})^6}}.$$

(8)

The condition $x = y$ represents mirror symmetry. It can be inferred that the Hamiltonian of the fourth-order Rubik's Cube matrix is solvable. **Fig. 3 (a)(b)(c)(d)** shows a 3D display of four feature roots.

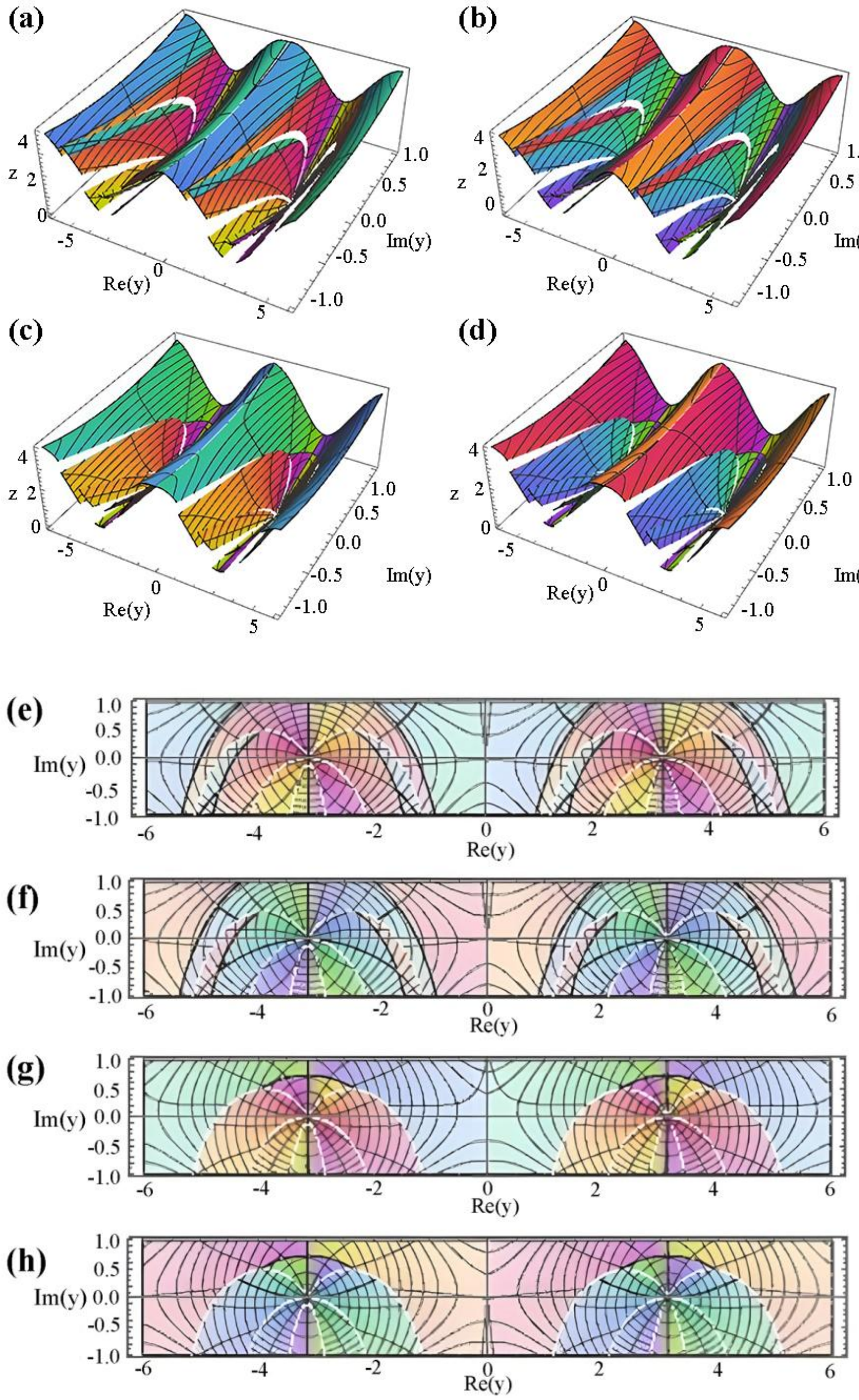

**FIG. 3 (a) (b) (c) (d)** 3D complex graphs of four characteristic roots, with the *z*-axis being the calculated complex solution. **(e) (f) (g) (h)** 2D angle plots of four feature roots. The colors in the image change with changes in the *x*-, *y*-, and *z*-axes.

There exists a base point (0, 3.1) in both real and virtual spaces, at which both points intersect, and the images at the left and right ends are similar, resulting in X-shaped Dirac points. In **Fig. 3 (a) (b) (c) (d),** the range of *Re*(*y*) is (-5-I,-5+I) , that of *Im*(*y*) is (-1,1) , and that of the complex solution is (0,4). In **Fig. 3 (e) (f) (g) (h)**, there is a straight line at *Re*(*y*)=3.1 that intersects three white parabolas at *Im(y)* = 0. The left branch of the first parabola has a maximum value of 0.5 on the imaginary axis and passes through the point (1.1, -1.0). The left branch of the second parabola has a maximum value of 0 and passes through the point (2.1, -1.0). The maximum of the left branch of the third parabola is the same as that of the second parabola; it passes through the point (2.8, -1.0). The right branch of the first parabola has a maximum value of 0.5 on the imaginary axis and passes through the point (5.1, -1.0); that of the second parabola passes through (4.1, -1.0), and that of the third passes through (3.4, -1.0). These six white parabolas are symmetric about *Re*(*y*)=0 on the negative half axis.

A phase of non-hermitian energy for non-abelian groups is displayed, which exhibits first symmetry in even space. As is well known, the general Hamiltonian only has real numbers, such as 1 Hermitian and 2 Hermitian, eliminating the imaginary part. In non-hermitian cases, only imaginary parts will be added to the energy band in the optical field, leading to band splitting and reflecting-band dispersion phenomena.

A numerical calculation with $\left(\gamma = 1, \lambda = 1, x = -y\right)$ leads to the eigenvalues:

$$
\begin{gathered}
-\mathrm{e}^{-\mathrm{i}x}\sqrt{1 + 2\mathrm{e}^{\mathrm{i}x} + 2\mathrm{e}^{2\mathrm{i}x} + 2\mathrm{e}^{3\mathrm{i}x} + \mathrm{e}^{4\mathrm{i}x} - \sqrt{(1 + \mathrm{e}^{\mathrm{i}x})^6(1 - 3\mathrm{e}^{\mathrm{i}x} + \mathrm{e}^{2\mathrm{i}x})}}, \\
\mathrm{e}^{-\mathrm{i}x}\sqrt{1 + 2\mathrm{e}^{\mathrm{i}x} + 2\mathrm{e}^{2\mathrm{i}x} + 2\mathrm{e}^{3\mathrm{i}x} + \mathrm{e}^{4\mathrm{i}x} - \sqrt{(1 + \mathrm{e}^{\mathrm{i}x})^6(1 - 3\mathrm{e}^{\mathrm{i}x} + \mathrm{e}^{2\mathrm{i}x})}}, \\
-\mathrm{e}^{-\mathrm{i}x}\sqrt{1 + 2\mathrm{e}^{\mathrm{i}x} + 2\mathrm{e}^{2\mathrm{i}x} + 2\mathrm{e}^{3\mathrm{i}x} + \mathrm{e}^{4\mathrm{i}x} + \sqrt{(1 + \mathrm{e}^{\mathrm{i}x})^6(1 - 3\mathrm{e}^{\mathrm{i}x} + \mathrm{e}^{2\mathrm{i}x})}}, \\
\mathrm{e}^{-\mathrm{i}x}\sqrt{1 + 2\mathrm{e}^{\mathrm{i}x} + 2\mathrm{e}^{2\mathrm{i}x} + 2\mathrm{e}^{3\mathrm{i}x} + \mathrm{e}^{4\mathrm{i}x} + \sqrt{(1 + \mathrm{e}^{\mathrm{i}x})^6(1 - 3\mathrm{e}^{\mathrm{i}x} + \mathrm{e}^{2\mathrm{i}x})}}.
\end{gathered}
\tag{9}
$$

Here, the imaginary number $i$ is time-dependent in phase, indicating the existence of time reversal symmetry when $x = -y$. It can be inferred that the Hamiltonian of the fourth-order quantum Rubik's Cube is solvable. **Fig. 4 (a)(b)(c)(d)** shows the four feature roots in 3D.

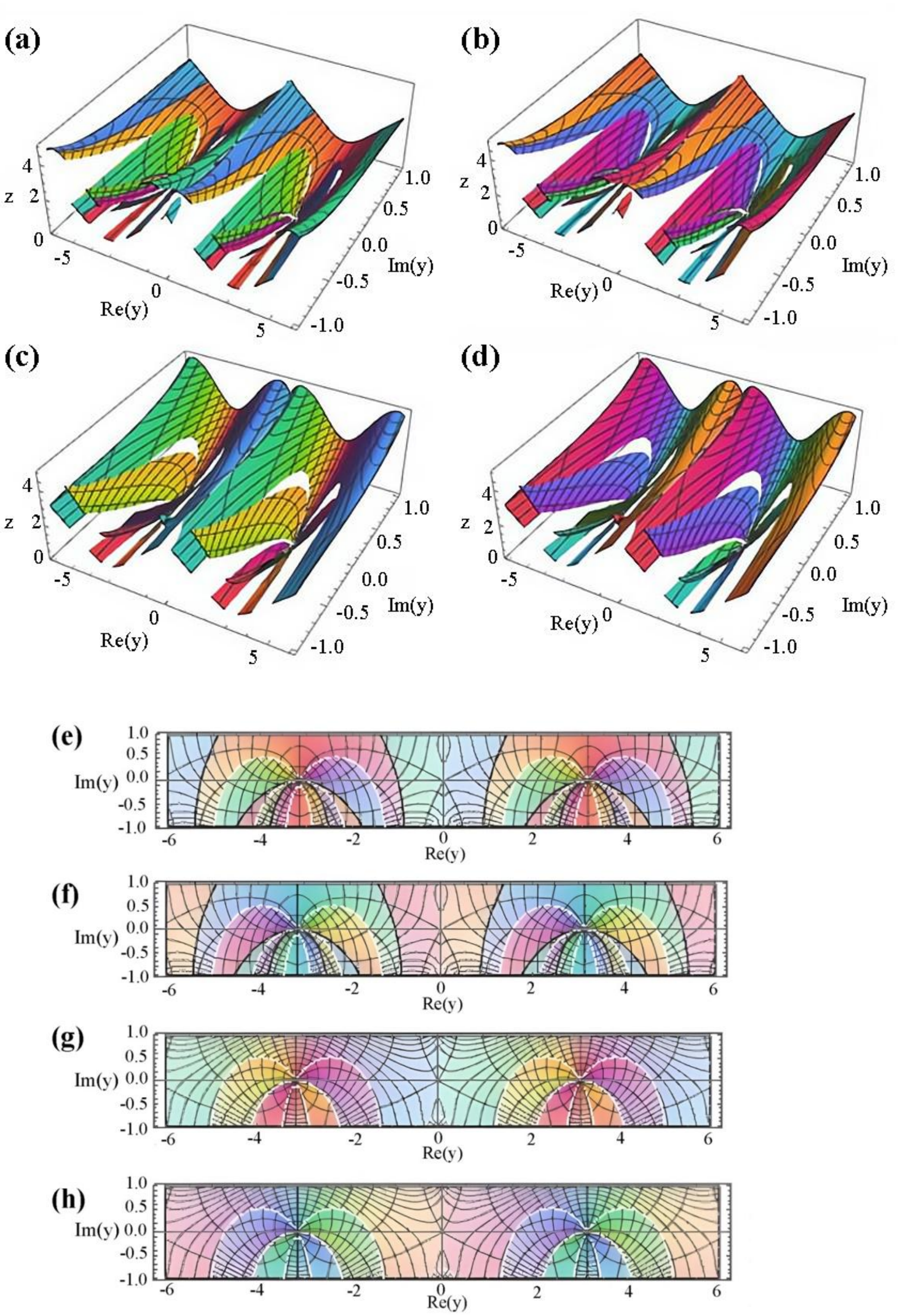

**FIG. 4 (a) (b) (c) (d)** 3D complex graphs of four characteristic roots, with the *z*-axis being the calculated complex solution. **(e) (f) (g) (h)** 2D angle plots of four feature roots. The colors in the image change with changes in the *x*-, *y*-, and *z*-axes.

There exists a base point (0,3.1) in both real and virtual spaces, at which both points intersect, and the images at the left and right ends are similar, resulting in an X-shaped Dirac point. The parabolas in **Fig. 4** are extremely similar to those in **Fig. 3**, although that the left branch of the first parabola passes through the point (1.2, -1.0).

### 2.2 Perturbation Theory and Quantum Rubik's Cube

Assume that an electric field $F$ is applied along the $z$ direction, the energy level of the $H$ atom is found, the eigenstates 1$s$, 2$s$, 2$p_x$, 2$p_y$ are used as the basis set, and the Hamiltonian matrix is recorded. If there are no fields, the matrix will be diagonal[28]:

$$\begin{array}{c} \\ [H_0]= \end{array}\begin{array}{c} \begin{matrix} 1s & 2s & 2p_x & 2p_y & 2p_z \end{matrix} \\ \begin{bmatrix} E_1 & 0 & 0 & 0 & 0 \\ 0 & E_2 & 0 & 0 & 0 \\ 0 & 0 & E_2 & 0 & 0 \\ 0 & 0 & 0 & E_2 & 0 \\ 0 & 0 & 0 & 0 & E_2 \end{bmatrix} \end{array},$$

(10)

where $E_0$ =13.6 eV, $E_1 = -E_0$, and $E_2 = -E_0/4$. The electric field leads to a matrix [$H_F$], which must be added to the matrix [$H_0$] . The $nm$-element of this matrix is

$$[H_F]_{nm} = qF\int_0^\infty drr^2\int_0^\pi \sin\theta d\theta\int_0^{2\pi} d\phi u_n^*(\vec{r})r\cos\theta u_m(\vec{r}).$$

(11)

Let $a_0$ be the Bohr radius and $q$ be the charge. Then, using the wave functions

$$\begin{aligned} u_{1s} &= \sqrt{1/\pi a_0^3}e^{-r/a_0} \\ u_{2s} &= \sqrt{1/32\pi a_0^3}(2-\frac{r}{a_0})e^{r/2a_0} \\ u_{2p_x} &= \sqrt{1/16\pi a_0^3}\left(\frac{r}{a_0}\right)e^{-r/2a_0}\sin\theta\cos\phi \\ u_{2p_y} &= \sqrt{1/16\pi a_0^3}\left(\frac{r}{a_0}\right)e^{-r/2a_0}\sin\theta\sin\phi \\ u_{2p_z} &= \sqrt{1/16\pi a_0^3}\left(\frac{r}{a_0}\right)e^{-r/2a_0}\cos\theta \end{aligned},$$

(12)

we can integrate and find that

$$[H_F] = \begin{array}{c} \\ \\ \\ \\ \\ \end{array}\begin{array}{ccccc} 1s & 2s & 2p_x & 2p_y & 2p_z \end{array}$$

$$[H_F] = \begin{bmatrix} 0 & 0 & 0 & 0 & A \\ 0 & 0 & 0 & 0 & B \\ 0 & 0 & 0 & 0 & 0 \\ 0 & 0 & 0 & 0 & 0 \\ A & B & 0 & 0 & 0 \end{bmatrix},$$

(13)

where A and B are linear functions of the domain:

$$A = \left(128\sqrt{2}/243\right) a_0 F, B = -3a_0 F.$$

(14)

Therefore,

$$\begin{array}{ccccc} 1s & 2s & 2p_z & 2p_x & 2p_y \end{array}$$

$$[H_0] + [H_F] = \begin{bmatrix} E_1 & 0 & A & 0 & 0 \\ 0 & E_2 & B & 0 & 0 \\ A & B & E_2 & 0 & 0 \\ 0 & 0 & 0 & E_2 & 0 \\ 0 & 0 & 0 & 0 & E_2 \end{bmatrix}.$$

(15)

The rows and columns have been re-arranged in Equation (15) to emphasize the fact that the $2p_x$ and $2p_y$ levels (unlike the $1s$, $2s$, and $2p_z$ levels) are decoupled from the rest of the matrix and are not affected by the fields.

We can simulate and implement this transformation using a fifth-order Rubik's Cube. The following demonstrates the operation of the fifth-order Rubik's Cube matrix and uses topology to restore the fifth-order Rubik's Cube, reducing it from three dimensions to two, as shown in **Fig. 2 (e)(f)(g)**. **Fig. 2 (f)** shows the topological structure of a fifth-order Rubik's Cube, consisting of three intersecting sets of five concentric circles each, with a total of six sets of intersection points. Each intersection point represents one of the six faces of a Rubik's Cube, with 5×5 circles on each face.

Here,

$$\Gamma_0=\begin{vmatrix} E_1 & & A & & \\ & E_2 & B & & \\ A & B & E_2 & & \\ & & & E_2 & \\ & & & & E_2 \end{vmatrix},\quad \Gamma_1=\begin{vmatrix} E_1 & & & & \\ & E_2 & & & \\ & & E_2 & & \\ & & & E_2 & \\ & & & & E_2 \end{vmatrix},\quad \Gamma_2=\begin{vmatrix} E_1 & & & & \\ & E_2 & & & \\ & & E_2 & & \\ & & & E_2 & \\ & & & & E_2 \end{vmatrix},$$

$$\Gamma_3=\begin{vmatrix} E_1 & & A & & \\ & E_2 & B & & \\ & & E_2 & & \\ & & & E_2 & \\ & & A & B & E_2 \end{vmatrix},\quad \Gamma_4=\begin{vmatrix} E_1 & & & & \\ & E_2 & & & \\ A & B & E_2 & & A \\ & & & E_2 & B \\ & & & & E_2 \end{vmatrix}, \text{ a } \quad \Gamma_5=\begin{vmatrix} E_1 & & & & \\ & E_2 & & & \\ & & E_2 & & \\ & & & E_2 & \\ & & & & E_2 \end{vmatrix}.$$

(16)

It can be shown that after rotation:

$$\begin{cases} [\![M]\!]_{2\times 2} \left| MB_{2(1)}D'_{1(2)}\, R_{1(3)}ML_{1(4)}MU_{1(5)} \right\rangle = [\![M']\!]_{2\times 2} \\ \qquad\qquad (\Gamma'_4) \\ [\![M']\!]_{5\times 5} = (\Gamma'_3)(\Gamma'_0)(\Gamma'_1)(\Gamma'_5) \\ \qquad\qquad (\Gamma'_2) \end{cases}.$$

(17)

According to degenerate perturbation theory, in the absence of a field, there are one eigenvalue $E_1$ and four degenerate eigenvalues $E_2$. When the difference between the off-diagonal term $H_{mn}$ and the corresponding diagonal term ($H_{mm}$-$H_{nn}$) is equal, the eigenvalue will not be equal to the diagonal value. This means that whenever there are two degenerate eigenvalues (i.e., $H_{mm}$ - $H_{nn}$=0), even a small non-diagonal element $H_{mn}$ will have a significant impact. Therefore, we anticipate that the electric field will have a notable impact on the initially degenerate 2*s* and $2p_z$ energy states. By focusing solely on the subset of the [*H*] matrix that encompasses these specific levels,

$$\begin{matrix} & & 2s & 2p_z \end{matrix}$$
$$[H_0]+[H_F]=\begin{bmatrix} E_2 & B \\ B & E_2 \end{bmatrix},$$

(18)

we can calculate the eigenvalues as $E = E_2 \pm B$, with the respective eigenvectors

$$|2s\rangle - |2p_z\rangle \text{ and } |2s\rangle + |2p_z\rangle.$$

(19)

This simplified method, known as degenerate perturbation theory, provides a good approximation of the exact eigenvalues, as illustrated in **Fig. 5**. This holds true when the off-diagonal elements (such as A) linking these levels to other states are significantly smaller than the energy discrepancies between them (such as $E_2 - E_1$). As shown in **Fig. 5**, the opening of the energy band (i.e., the rate of change) will increase with the increase of $a_0$. The solid line in the figure represents the result obtained through direct diagonalization, while the data points represent the perturbation-theory results $E = E_2 \pm B$.

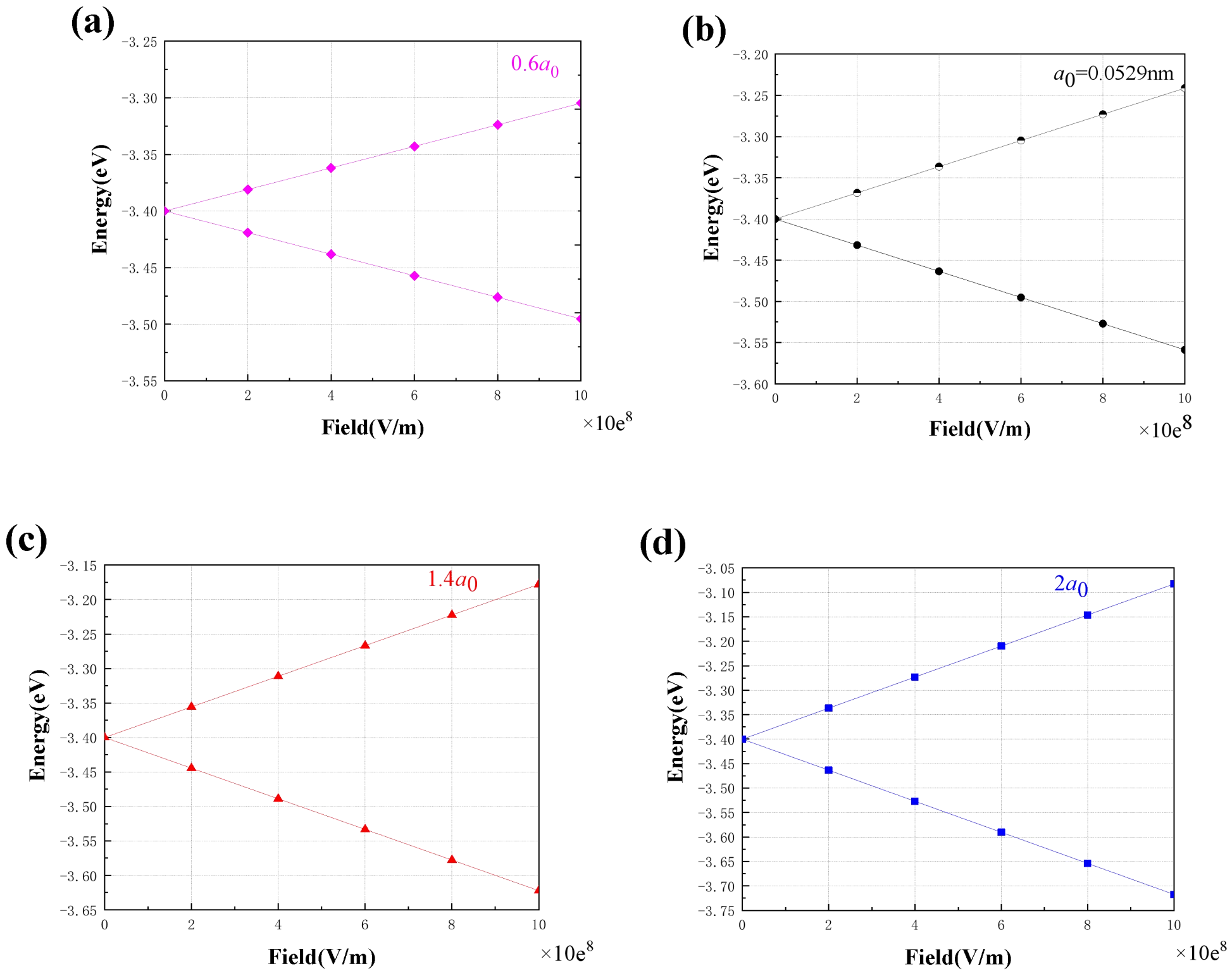

**FIG. 5** The difference in energy between $2s$ and $2p_z$ levels when atomic radius is taken as (a) $0.6a_0$, (b) $a_0$, (c) $1.4a_0$, and (d) $2a_0$.

## 3. Conclusions

The quantum Rubik's Cube in this study is linked to the BBH model, resulting in a new Hamiltonian mathematical form. The matrix algorithm for the quantum Rubik's Cube is defined based on convolution, and the calculation results show that the resulting Hamiltonian is solvable. In order to obtain the operation process of the matrix of the quantum Rubik's Cube from path integration, a two-dimensional structure-topology diagram of the quantum Rubik's Cube expansion is drawn using a Josephus ring. In addition, the dynamic simulation of electronic orbital transitions using a fifth-order quantum Rubik's Cube is investigated on the basis of perturbation theory; this approximate method provides accurate energy values for energy levels. This study has brought new ideas and methods to the field of quantum computing.

# Support Information

## Understanding Energy-Level Structure Using a Quantum Rubik's Cube

Yu Wang, Maolin Bo*

Key Laboratory of Extraordinary Bond Engineering and Advanced Materials Technology (EBEAM) of Chongqing, Yangtze Normal University, Chongqing 408100, China

*Corresponding author: Maolin Bo (E-mail addresses: bmlwd@yznu.edu.cn)

### 1.THE OPERATION OF QUANTUM RUBIK'S CUBE

The matrix size of the quantum Rubik's Cube can be determined through convolution. Matrix convolution is a mathematical transformation from one-dimensional to two-dimensional convolution, written as:

$$O_{m,n} = f\left(\sum_{u=1}^{U}\sum_{v=1}^{V} w_{u,v} \cdot x_{i+u,j+v} + b\right). \tag{S1}$$

In the formula, $w_{u,v}$ represents the weight value at row *u* and column *v* of the convolution kernel; $x_{i+u,j+v}$ represents the value located in the $i+u$ row and $j+v$ column of the image or feature map; *i* and *j* respectively indicate that the convolution kernel slides *i* rows and *j* columns on the matrix or feature map; *b* is the bias term; and $f(\bullet)$ is an activation function that can increase the nonlinearity of the network.

Fully connected layers (FCs) play the role of classifiers in the entire convolutional neural network. Each node in the FC is connected to all nodes in the previous layer and is used to integrate the features extracted from the previous layer. Assume that there are *n* channels in the *I*-th layer (which means that there are *n* feature maps) and *m* features in the *I+1*-st layer. Each channel map in the *I*-th layer has its corresponding error-sensitivity value, and its calculation is based on the contributions of all feature kernels in the *I+1*-st layer. The error-sensitivity value of the *j*-th check in layer I+1 and the *i*-th channel of layer *I* is calculated from

$$\delta_i^I = \sum_{j=1}^{m} \delta_i^{I+1} \otimes w_{ij}$$

(S2)

In the formula, $\otimes$ represents the convolution operation of the matrix. When calculating the error sensitivity of the *i*-th channel, all the kernels of layer *I+2* must be calculated and summed up. Moreover, because the default pooling layer here is a linear activation function, the derivative of the corresponding node is not multiplied later.

When a 3×3 feature map passes through a 2×2 feature map with a step size of 1, its value is given by

$$\begin{bmatrix} a_{11} & a_{12} & a_{13} \\ a_{21} & a_{22} & a_{23} \\ a_{31} & a_{32} & a_{33} \end{bmatrix} * \begin{bmatrix} w_{11} & w_{12} \\ w_{21} & w_{22} \end{bmatrix} = \begin{bmatrix} z_{11} & z_{12} \\ z_{21} & z_{22} \end{bmatrix}.$$

(S3)

By the definition of the convolution operation, this means that

$$\begin{cases} z_{11} = a_{11}w_{11} + a_{12}w_{12} + a_{21}w_{21} + a_{22}w_{22} \\ z_{12} = a_{12}w_{11} + a_{13}w_{12} + a_{22}w_{21} + a_{23}w_{22} \\ z_{21} = a_{21}w_{11} + a_{22}w_{12} + a_{31}w_{21} + a_{32}w_{22} \\ z_{22} = a_{22}w_{11} + a_{23}w_{12} + a_{32}w_{21} + a_{33}w_{22} \end{cases}.$$

(S4)

Here, $a_{11}$ is the coefficient only of $w_{11}$; $a_{12}$ is the coefficient of both $w_{12}$ and $w_{11}$; and so on. In the reverse process, assuming errors $\delta_{11}, \delta_{12}, \delta_{21}, \delta_{22}$ to form a 2×2 matrix, mark $\nabla a$ as the reverse gradient. Inverting Eq. (S4), we obtain:

$$\begin{cases} \nabla a_{11} = \delta_{11}w_{11} \\ \nabla a_{12} = \delta_{12}w_{12} + \delta_{12}w_{11} \\ \nabla a_{13} = \delta_{12}w_{12} \\ \nabla a_{21} = \delta_{11}w_{21} + \delta_{21}w_{11} \\ \nabla a_{22} = \delta_{11}w_{22} + \delta_{12}w_{21} + \delta_{21}w_{12} + \delta_{22}w_{11} \\ \nabla a_{23} = \delta_{12}w_{22} + \delta_{22}w_{12} \\ \nabla a_{31} = \delta_{21}w_{21} \\ \nabla a_{32} = \delta_{21}w_{22} + \delta_{22}w_{21} \\ \nabla a_{33} = \delta_{22}w_{22} \end{cases}.$$

(S5)

These nine equations can written more simply using matrix convolution:

$$\begin{bmatrix} 0 & 0 & 0 & 0 \\ 0 & \delta_{11} & \delta_{12} & 0 \\ 0 & \delta_{21} & \delta_{22} & 0 \\ 0 & 0 & 0 & 0 \end{bmatrix} * \begin{bmatrix} w_{22} & w_{21} \\ w_{12} & w_{11} \end{bmatrix} = \begin{bmatrix} \nabla a_{11} & \nabla a_{12} & \nabla a_{13} \\ \nabla a_{21} & \nabla a_{22} & \nabla a_{23} \\ \nabla a_{31} & \nabla a_{32} & \nabla a_{33} \end{bmatrix} .$$

(S6)

The convolution kernel of backward propagation is obtained by rotating that of forward propagation by 180°. Taking two convolution kernels as an example of the multi-kernel case and assuming a channel graph size of 3×3 in the first layer and two 2×2 convolution kernels in the first +1 layer, during forward propagation convolution the *I*+*1*-st layer will have two 2×2 convolution kernels, as shown in **Fig. S1**.

| | |
|---|---|
| 0.4 | 0.2 |
| 0.2 | 0.1 |
| 0.2 | 0.1 |
| 0.1 | -0.3 |

**FIG. S1** Two convolution kernels of size 2×2.

When using backpropagation to calculate the error sensitivity, assume that the error-sensitivity values of the two convolution kernels in the *I*+1-st layer are already known, as shown in **Fig. S2**. They can then be filled with a filling function to obtain **Fig. S3**.

| | |
|---|---|
| 1 | 3 |
| 2 | 2 |
| 2 | 1 |
| 1 | 1 |

**FIG. S2** Error-sensitivity values of convolutional kernels

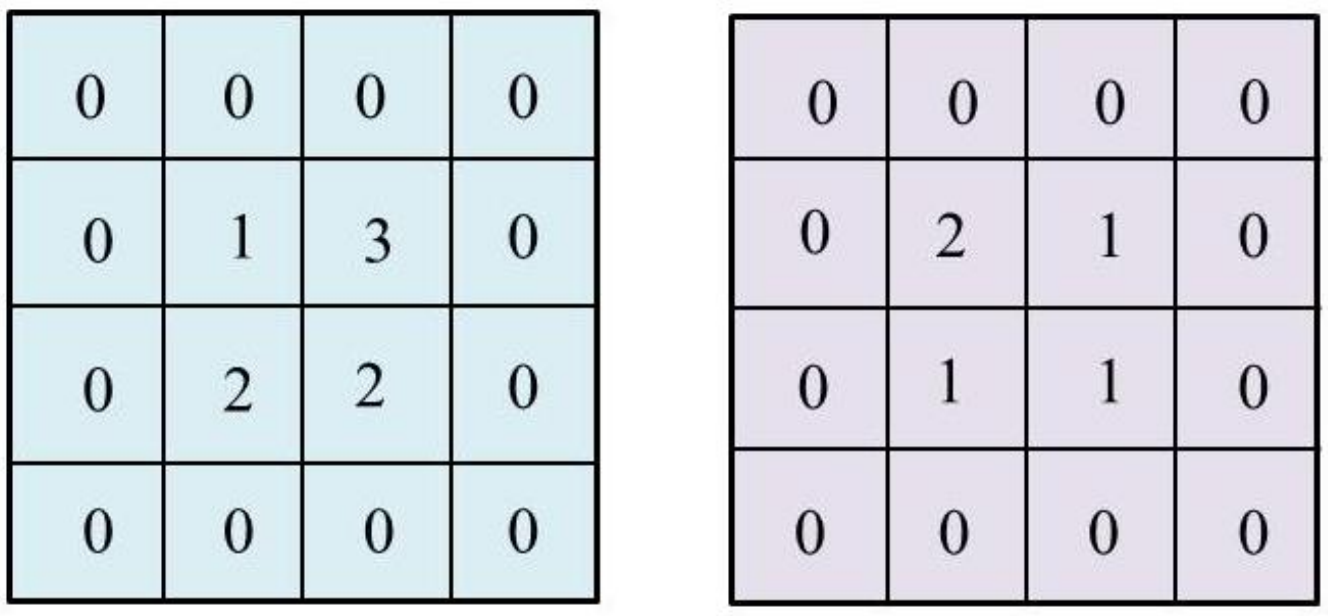

**FIG. S3** The 2×2 matrices in Fig. S2 expanded to size 4×4 with an all-zero filling function.

The convolution operation is shown in **Fig. S4**.

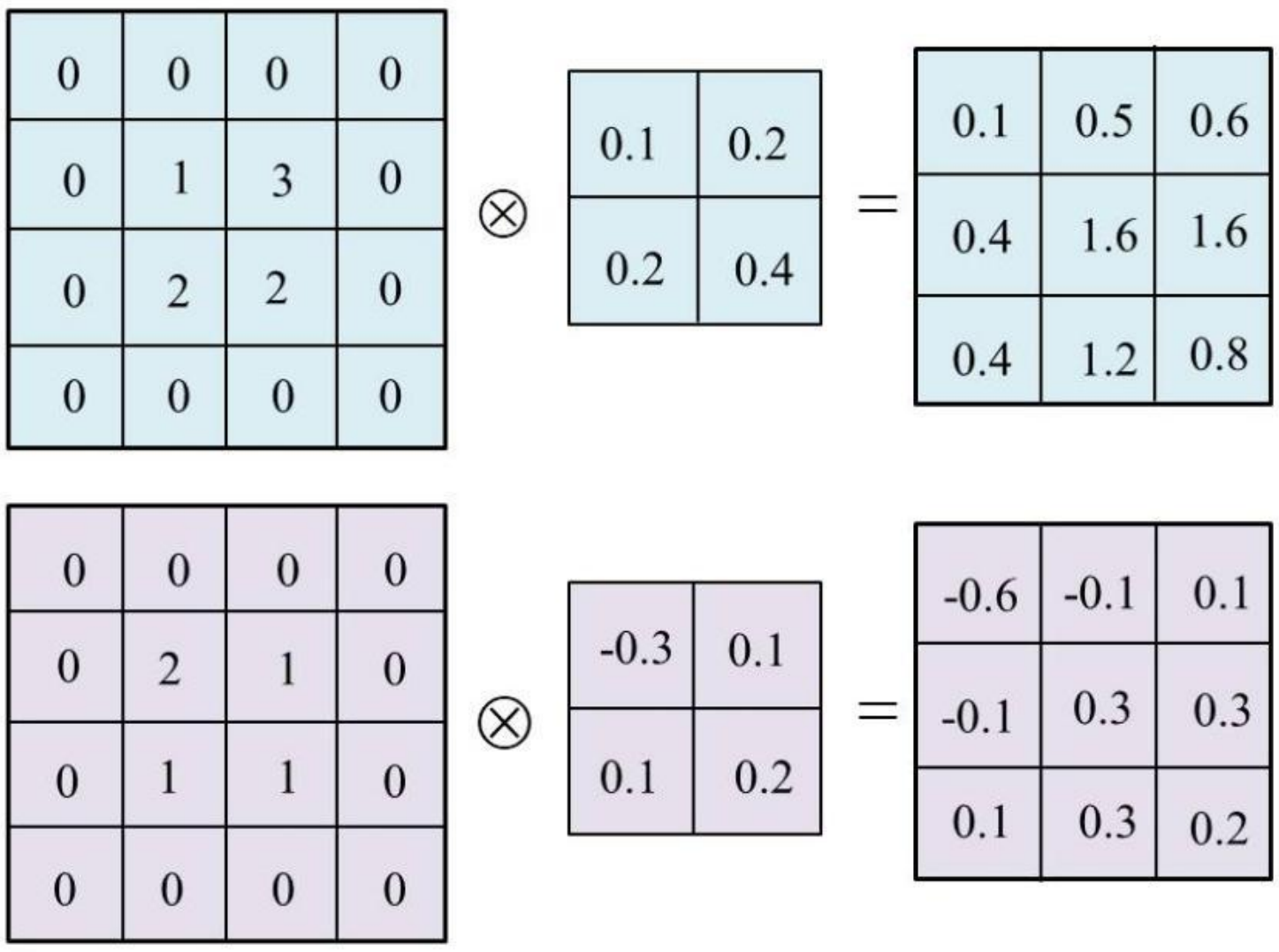

**FIG. S4** Convolutional process.

Finally, the convolution results are added together to obtain **Fig. S5**.

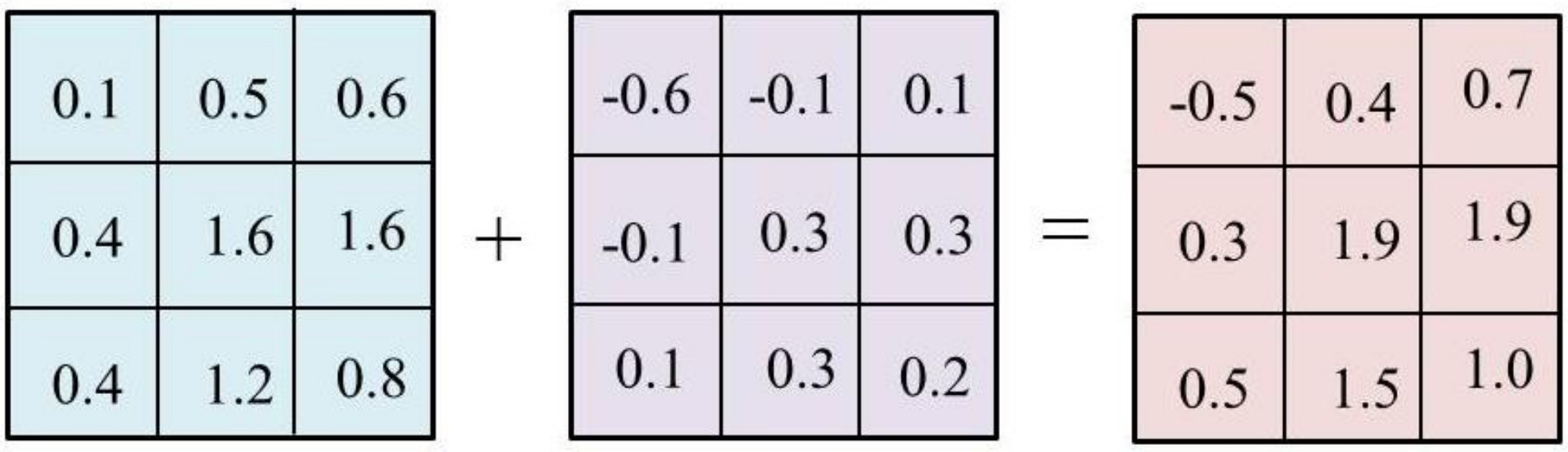

**FIG. S5** Adding convolution results of Fig. S4.

At this point, the sensitivity value of backpropagation error has been calculated. The above quantum Rubik's Cube convolution operation is defined by the properties of the quantum Rubik's Cube group $\mathfrak{R}$. The convolution operation can change the size of the quantum Rubik's Cube matrix. The quantum Rubik's Cube convolution operation provides possibilities for quantum computing in neural networks. In summary, the matrix operation rules of quantum Rubik's Cube satisfy the allocation rate, exchange rate, and associative law of addition.

## 2. MATRIX AND PATH INTEGRATION OF QUANTUM RUBIK'S CUBE

The Feynman path-integral method is an effective tool for solving quantum mechanics problems. It considers particle motion from the perspective of the entire path and obtains the probability amplitude of particles at different positions by summing up the phase factors of all possible paths. Although its mathematical form is relatively complicated, the Feynman path-integral method provides an intuitive perspective on the behavior and interactions of particles, providing important tools for studying the physical processes of the microscopic world.

The propagator *K*(*b,a*) is defined as

$$\psi\left(x_b,t_b\right)=\int_{-\infty}^{\infty} dx_a K\left(x_b,t_b;x_a,t_a\right)\psi\left(x_a,t_a\right) \qquad \left(t_b \geq t_a\right),$$

(S7)

which is usually written more compactly as

$$\psi\left(x_b,t_b\right)=\int dx_a K\left(b,a\right)\psi\left(x_a,t_a\right) \qquad \left(t_b \geq t_a\right).$$

(S8)

If the propagator *K* (*b, a*) is known, the wave function $\psi\left(x_b,t_b\right)$ at any subsequent time can be calculated from the earlier wave function $\psi\left(x_a,t_a\right)$, and the properties of the quantum system can be determined accordingly.

If $\psi\left(x_a,t_a\right)=\delta\left(x_a-x_0\right)$ is set, then

$$\psi\left(x_b,t_b\right)=\int dx_a K\left(b,a\right)\delta\left(x_a-x_0\right)=K\left(x_b,t_b;x_0,t_a\right).$$

(S9)

Therefore, if the particle is located at $x_a$ at time $t_a$, the propagator is the probability-wave amplitude (wave function) of finding the particle transmitted from $\left(x_a,t_a\right)$ at time $t_b$ and $x_b$. The path of the propagator is shown in **Fig. S6 (a)**.

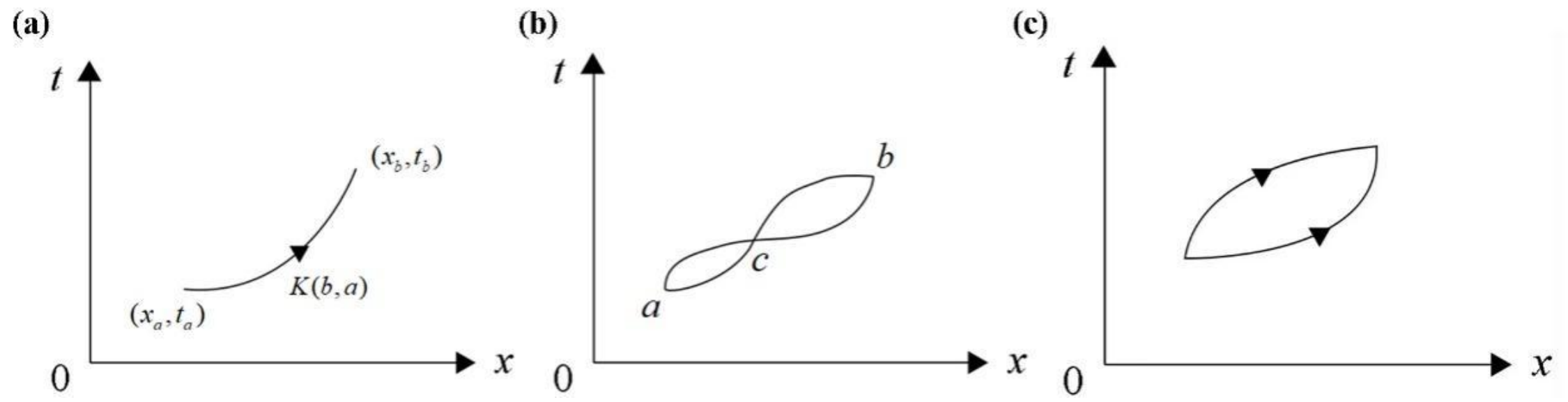

**FIG. S6** Schematic diagram of path integration

The propagator is a special wave function that represents the influence of a point source. It is the Green's function of the Schrödinger equation. According to the definition of propagator:

$$\psi\left(x_b,t_b\right)=\int dx_c K(b,c)\psi(x_c,t_c) \qquad \left(t_b \geq t_c\right)$$
$$\psi\left(x_c,t_c\right)=\int dx_a K(c,a)\psi(x_a,t_a) \qquad \left(t_c \geq t_a\right).$$

(S10)

In **Fig. S6 (b)**, the coordinates of points *a*, *b*, and *c* correspond to those in **Eq. S10**. Therefore, $\psi\left(x_b,t_b\right)=\int dx_a K(b,a)\psi(x_a,t_a)$. By comparing that Equ. S10, we can see the transitivity of propagators:

$$K(b,a)=\int dx_c K(b,c)K(c,a) \qquad \left(t_b \geq t_c \geq t_a\right).$$

(S11)

To prove the normalization of propagators

$$\psi(x_b,t_b)=\int dx_a K(b,a)\psi(x_a,t_a)$$

(S12)

When $t_b = t_a$, compare

$$\psi(x_b,t_b)=\int dx_a \delta(x_b - t_a)\psi(x_a,t_a)$$

(S13)

It is known that $\lim_{t_b \to t_a} K(b,a)=\delta(x_b - x_a)$. According to Feynman's hypothesis, the path of particle propagation from point $(x_a, t_a)$ to point $(x_b, t_b)$ can be arbitrary, but each path appears with a different probability. **Fig. S6 (c)** shows the propagation paths of two particles.

The probability amplitude that the path appears is proportional to $e^{\frac{i}{\hbar}S}$ ($S$ being the amount of action of the path); the propagator is

$$K(b,a) = C \sum_{all-paths} e^{\frac{i}{\hbar}S} = \int_a^b e^{\frac{i}{\hbar}S} Dx(t) \quad \text{(S14)}$$

In the formula, $C$ is the normalization coefficient, and $\int Dx(t)$ represents the integration of all possible paths from the initial point $x(t_a) = x_a$ to the termination point $x(t_b) = x_b$.

The path integral of propagator $K(b,a) = \int_a^b e^{\frac{i}{\hbar}S} Dx(t)$ is a functional integral, which is generally difficult to calculate. Feynman proposed a computational scheme for polygonal paths, dividing the functional product into a series of definite integrals for calculation. The transitivity of propagators is

$$\begin{aligned} K_N(b,a) &= \int_{-\infty}^{\infty} dx_1 \int_{-\infty}^{\infty} dx_2 \cdots \int_{-\infty}^{\infty} dx_{N-1} K(b, N-1) \cdots K(j+1, j) \cdots K(2,1) K(1,a) \\ &= \int dx_1 \int dx_2 \cdots \int dx_{N-1} \prod_{j=0}^{N-1} K(j+1, j) \end{aligned} \quad \text{(S15)}$$

When $N$ is large enough, the close proximity between points $j$ and $j+1$ allows $K(j+1, j)$ to be written as

$$K(j+1, j) = \frac{1}{A} e^{\frac{i}{\hbar}S(j+1,j)},$$

and therefore

$$K_N(b,a) = \frac{1}{A^N} \int dx_1 \int dx_2 \cdots \int dx_{N-1} \exp\left(\frac{i}{\hbar} S_N\right), \quad \text{(S16)}$$

where A is the normalization factor. For one-dimensional free particles, using Fresnel integration and substituting into $N_\varepsilon = t_b - t_a, x_0 = x_a, x_N = x_b$ , the propagator of one-dimensional free particles can be obtained:

$$K(b,a) = \left[\frac{m}{2\pi i\hbar(t_b - t_a)}\right]^{\frac{1}{2}} \exp\left(\frac{im(x_b - x_a)^2}{2\hbar(t_b - t_a)}\right).$$

(S17)

According to Feynman's path-integral theory, the state $\psi(x, t+\varepsilon)$ of the particle at time $t+\varepsilon\ (\varepsilon \to 0^+)$ is related to the state $\psi(x,t)$ of the particle at time *t* by

$$\psi(x,t+\varepsilon) = \int_{-\infty}^{\infty} dx' K(x,t+\varepsilon; x',t)\psi(x',t),$$

(S18)

and $K(x,t+\varepsilon;x',t) = \frac{1}{A}\exp\left[\frac{i\varepsilon}{\hbar}L\left(\frac{x+x'}{2},\frac{x-x'}{\varepsilon},t\right)\right]$ . If a particle moves in a potential field $V(x,t), L = \frac{m}{2}\dot{x}^2 - V(x,t)$, then

$$\psi(x,t+\varepsilon) = \frac{1}{A}\int_{-\infty}^{\infty} dx'\psi(x',t)\exp\left\{\frac{i\varepsilon}{\hbar}\left[\frac{m}{2}\left(\frac{x+x'}{2}\right)^2 - V\left(\frac{x-x'}{\varepsilon},t\right)\right]\right\}.$$

(S19)

Using the integral formula $\int_{-\infty}^{\infty} e^{i\alpha\xi^2} d\xi = \sqrt{\frac{i\pi}{\alpha}}$ , it follows that

$$i\hbar\frac{\partial\psi}{\partial t} = -\frac{\hbar^2}{2m}\frac{\partial^2\psi}{\partial x^2} + V\psi.$$

(S20)

This is exactly the one-dimensional Schrödinger equation: Feynman path-integral theory is completely equivalent to Schrödinger wave mechanics. Therefore, we can also discuss propagators starting from the Schrödinger equation

$$i\hbar\frac{\partial}{\partial t}\left|\psi(t)\right\rangle = H\left|\psi(t)\right\rangle .$$

(S21)

If the Hamiltonian $H$ does not depend explicitly on time, then

$$\left|\psi(t_b)\right\rangle = e^{-\frac{i}{\hbar}H(t_b-t_a)}\left|\psi(t_a)\right\rangle .$$

(S22)

In the coordinate representation,

$$\left\langle x_b \middle| \psi(t_b)\right\rangle = \int dx_a \left\langle x_b \right| e^{-\frac{i}{\hbar}H(t_b-t_a)}\left|x_a\right\rangle\left\langle x_a \middle| \psi(t_a)\right\rangle .$$

(S23)

By comparing $\psi(x_b,t_b) = \int dx_a K(b,a)\psi(x_a,t_a)$, we can obtain $\left\langle T_{ba}\right| = K(b,a) = \left\langle x_b \right| e^{-\frac{i}{\hbar}H(t_b-t_a)}\left|x_a\right\rangle$, while also knowing $\left|T_{ba}\right\rangle = \left\langle T_{ab}\right|$.

In the energy representation $H\left|n\right\rangle = E_n\left|n\right\rangle$,

$$\begin{aligned}K(b,a) &= \sum_n \left\langle x_b \right| e^{-\frac{i}{\hbar}H(t_b-t_a)}\left|n\right\rangle\left\langle n \middle| x_a\right\rangle \\ &= \sum_n e^{-\frac{i}{\hbar}E_n(t_b-t_a)}\left\langle x_b \middle| n\right\rangle\left\langle n \middle| x_a\right\rangle \\ &= \sum_n \psi_n(x_b)\psi_n^*(x_a)e^{-\frac{i}{\hbar}E_n(t_b-t_a)}\end{aligned}$$

(S24)

If the Hamiltonian $H$ (or Lagrangian $L$) does not explicitly contain time, then $K(b,a)$ is only a function of the time interval $(t_b - t_a)$. The quantum Rubik's Cube operator is defined as $K(b,a) = \left|T_{ab}\right\rangle := \left|\Re\right\rangle$, which has non-abelian group properties and can be represented by a path integral. $\left|\Re\right\rangle$ is a quantum Rubik's Cube matrix operator, and $\Re$ is the group of the quantum Rubik's Cube. Therefore, it can be concluded that

$$H|\Re\rangle = \pm\sum_{j,k} J_{jk} [\![M_j]\!]_{i\times i} |\Re\rangle = \pm\sum_{j,k} J_{jk} [\![M'_j]\!]_{i\times i}$$

.

(S25)

## 3. ISING MODEL OF QUANTUM RUBIK'S CUBE

The interaction between adjacent atoms in Hamiltonian can be expressed as

$$H = -J_{ij} \vec{S}_i \bullet \vec{S}_j$$

(S26)

where $J_{ij}$ is the coupling strength.

$$\hat{S}_x = \frac{\hbar}{2}\begin{pmatrix} 0 & 1 \\ 1 & 0 \end{pmatrix},\ \hat{S}_y = \frac{\hbar}{2}\begin{pmatrix} 0 & -i \\ i & 0 \end{pmatrix},\ \hat{S}_z = \frac{\hbar}{2}\begin{pmatrix} 1 & 0 \\ 0 & -1 \end{pmatrix}$$

(S27)

Using the relationship between $\hat{S}$ and $\sigma$, the component matrices, namely Pauli matrices, are obtained:

$$\sigma_x = \begin{pmatrix} 0 & 1 \\ 1 & 0 \end{pmatrix},\ \sigma_y = \begin{pmatrix} 0 & -i \\ i & 0 \end{pmatrix},\ \sigma_z = \begin{pmatrix} 1 & 0 \\ 0 & -1 \end{pmatrix}.$$

(S28)

The spin expansion of the magic cube matrix can be written as:

$$\begin{matrix} & \begin{pmatrix} -i & -i \\ 1 & -1 \end{pmatrix} & & \\ [\![M_c]\!]_{2\times 2} = \begin{pmatrix} 0 & 0 \\ 0 & 0 \end{pmatrix} & \begin{pmatrix} 1 & 1 \\ 1 & 1 \end{pmatrix} & \begin{pmatrix} 0 & 0 \\ 0 & 0 \end{pmatrix} & \begin{pmatrix} 0 & 0 \\ 0 & 0 \end{pmatrix} \\ & \begin{pmatrix} -1 & 1 \\ i & i \end{pmatrix} & & \end{matrix}$$

(S29)

The spin expansion of the magic cube matrix can get the form of the Pauli matrices of the magic cube by

rotating:

$$
\llbracket M_c \rrbracket_{2\times 2} = \begin{matrix} & \begin{pmatrix} -i & -i \\ 1 & -1 \end{pmatrix} & & \\ \begin{pmatrix} 0 & 0 \\ 0 & 0 \end{pmatrix} & \begin{pmatrix} 1 & 1 \\ 1 & 1 \end{pmatrix} & \begin{pmatrix} 0 & 0 \\ 0 & 0 \end{pmatrix} & \begin{pmatrix} 0 & 0 \\ 0 & 0 \end{pmatrix} \\ & \begin{pmatrix} -1 & 1 \\ i & i \end{pmatrix} & & \end{matrix} \Rightarrow \llbracket \sigma_x \rrbracket_{2\times 2} = \begin{matrix} & \begin{pmatrix} 1 & 0 \\ 0 & -1 \end{pmatrix} & & \\ \begin{pmatrix} 0 & i \\ -i & 0 \end{pmatrix} & \begin{pmatrix} 0 & 1 \\ 1 & 0 \end{pmatrix} & \begin{pmatrix} 0 & -i \\ i & 0 \end{pmatrix} & \begin{pmatrix} 0 & 1 \\ 1 & 0 \end{pmatrix} \\ & \begin{pmatrix} 1 & 0 \\ 0 & -1 \end{pmatrix} & & \end{matrix}
$$

$$
\llbracket M_c \rrbracket_{2\times 2} \left| D'_{1(1)} B'_{2(2)} D'_{2(3)} B'_{1(4)} D'_{2(5)} B'_{2(6)} L'_{2(7)} B'_{1(8)} \right\rangle = \llbracket \sigma_x \rrbracket_{2\times 2}
$$

(S30)

$$
\llbracket M_c \rrbracket_{2\times 2} = \begin{matrix} & \begin{pmatrix} -i & -i \\ 1 & -1 \end{pmatrix} & & \\ \begin{pmatrix} 0 & 0 \\ 0 & 0 \end{pmatrix} & \begin{pmatrix} 1 & 1 \\ 1 & 1 \end{pmatrix} & \begin{pmatrix} 0 & 0 \\ 0 & 0 \end{pmatrix} & \begin{pmatrix} 0 & 0 \\ 0 & 0 \end{pmatrix} \\ & \begin{pmatrix} -1 & 1 \\ i & i \end{pmatrix} & & \end{matrix} \Rightarrow \llbracket \sigma_y \rrbracket_{2\times 2} = \begin{matrix} & \begin{pmatrix} 0 & 1 \\ -1 & 0 \end{pmatrix} & & \\ \begin{pmatrix} 0 & 1 \\ 1 & 0 \end{pmatrix} & \begin{pmatrix} 0 & -i \\ i & 0 \end{pmatrix} & \begin{pmatrix} 0 & 1 \\ 1 & 0 \end{pmatrix} & \begin{pmatrix} 0 & i \\ -i & 0 \end{pmatrix} \\ & \begin{pmatrix} 0 & -1 \\ 1 & 0 \end{pmatrix} & & \end{matrix}
$$

$$
\llbracket \sigma_x \rrbracket_{2\times 2} \left| k_{x1} \right\rangle = \llbracket \sigma_y \rrbracket_{2\times 2}
$$

(S31)

$$
\llbracket M_c \rrbracket_{2\times 2} = \begin{matrix} & \begin{pmatrix} -i & -i \\ 1 & -1 \end{pmatrix} & & \\ \begin{pmatrix} 0 & 0 \\ 0 & 0 \end{pmatrix} & \begin{pmatrix} 1 & 1 \\ 1 & 1 \end{pmatrix} & \begin{pmatrix} 0 & 0 \\ 0 & 0 \end{pmatrix} & \begin{pmatrix} 0 & 0 \\ 0 & 0 \end{pmatrix} \\ & \begin{pmatrix} -1 & 1 \\ i & i \end{pmatrix} & & \end{matrix} \Rightarrow \llbracket \sigma_z \rrbracket_{2\times 2} = \begin{matrix} & \begin{pmatrix} 0 & 1 \\ 1 & 0 \end{pmatrix} & & \\ \begin{pmatrix} i & 0 \\ 0 & -i \end{pmatrix} & \begin{pmatrix} 1 & 0 \\ 0 & -1 \end{pmatrix} & \begin{pmatrix} i & 0 \\ 0 & -i \end{pmatrix} & \begin{pmatrix} -1 & 0 \\ 0 & 1 \end{pmatrix} \\ & \begin{pmatrix} 0 & 1 \\ 1 & 0 \end{pmatrix} & & \end{matrix}
$$

$$
\llbracket \sigma_x \rrbracket_{2\times 2} \left| k_{y1} \right\rangle = \llbracket \sigma_y \rrbracket_{2\times 2}
$$

(S32)

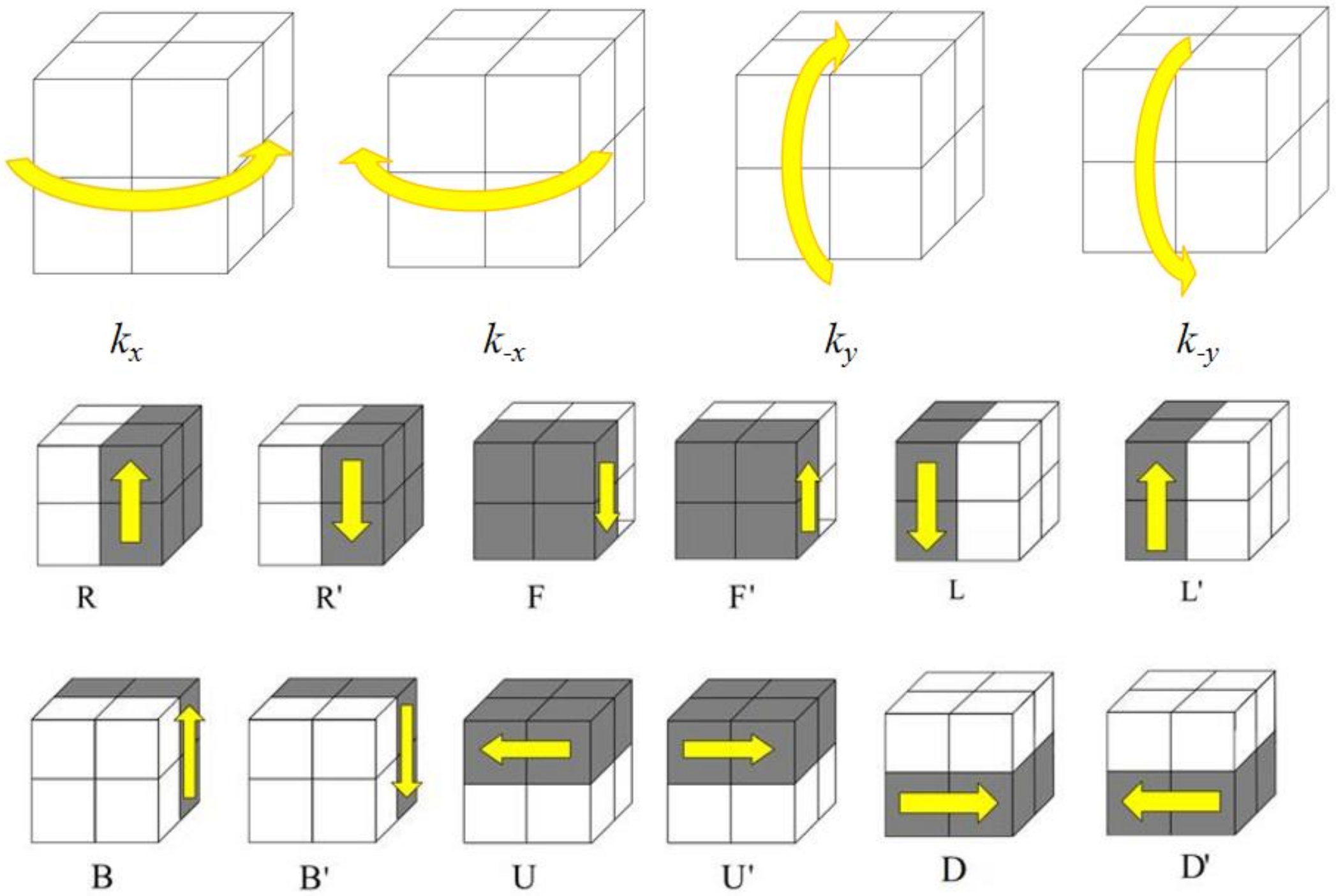

**FIG. S7** Rules for operating the quantum Rubik's cube matrix